# Geometric Phase in Kitaev Quantum Spin Liquid

Zheng-Chuan Wang

School of Physical Sciences,

University of Chinese Academy of Sciences, Beijing 100049, China.

wangzc@ucas.ac.cn

## Abstract

Quantum spin liquid has massive many spin entanglement in the ground state, we can evaluate it by the entanglement entropy, but the latter can not be observed directly by experiment. In this manuscript, we try to characterize it's topological properties by the geometric phase. However the usual adiabatic or non-adiabatic geometric phase can not appear in the density matrix of entanglement entropy, so we extend it to the sub-geometric phase which can exist in the density matrix and have influence on the entanglement entropy, spin correlation function as well as other physical observable. We will demonstrate that the imaginary part of sub-geometric phase will deviate the resonance peak by an amount concerning with this phase and affect the energy level crossing, while the real part of sub-geometric phase will determine the stability of initial state, it may provide a complement on the selection rule of quantum transition.

**PACS**: 03.65.Vf, 75.10.Kt, 75.10.Jm, 42.50.Lc

## I. Introduction

In recent years, quantum spin liquid (QSL) had attracted more and more attention, in which the extotic phase, fractionalized excitation and topological orders will occur[1]. QSL usually appear in the spin systems with frustration, i.e., the geometric frustration in the triangular lattice and Kagome lattice spin system[2], in which strong frustration leads to highly degenerate ground states and the long range magnetic ordering is suppressed by the corresponding fluctuation. Generally speaking, if the temperature of the frustrated system $T_c \leq T \leq |Q_{cw}|$, where $T_c$ is the anomalously low ordering temperature, and $Q_{cw}$ is the Curie-Weiss temperature, the system is remain in the regime of QSL[3]. Beside the geometric frustration, Kitaev proposed a model for QSL with anistropic exchange frustration[4], there exist the emergent excitation of Majorana Fermion in this strongly correlated system of quantum many body, which is topologically robust and can be used to carry qubit in quantum computation. Till now, Kitaev model had been experimental simulated by means of cold atom system[5], possible Kitaev materials include the honeycomb iridate material[6], i.e., $Na_2IrO_3$ and $Li_2IrO_3$, and the 4d-transition metal material α-$RuCl_3$[7], but they are still not the ideal QSL now, maybe they are close to QSL state or can be driven into a real QSL in some way[6].

The ground state of QSL has massive many-body entanglement, which can induce non-local topological excitation, how to characterize these topological properties by a proper physical quantity is not easy. Maybe one can use Von-Neumann entanglement entropy to evaluate the massive entanglement of QSL, but it can not be directly observed by experiment, we need find other physical observable. In this manuscript, we try to use geometric phase to characterize the topological properties of QSL, because this topological phase had been observed in many systems[8]. The Berry curvature in Kitaev honeycomb model had been studied by Bascone et al. with respect to Hamiltonian parameters[9], they explored the critical properties of Berry curvature during the topological phase transitions of this model, the mean Uhlmann curvature and the Uhlmann number were also investigated for the finite-temperature Kitaev model[10]. However, it should be pointed out that the usual geometric phase can not appear in the density matrix $\hat{\rho} = |\psi(\mathrm{t})><\psi(\mathrm{t})|$ of entanglement entropy, because the geometric phase $e^{i\beta(t)}$ in $|\psi(\mathrm{t})>$ will cancel with the phase $e^{-i\beta(t)}$ in it's conjugate state $<\psi(\mathrm{t})|$, although the adiabatic Berry phase had been extended to the non-cyclic and non-adiabatic case[11,12], they have no influence on the entanglement entropy at all. In 2013, Wang proposed a sub-geometric phase for the density matrix[13], which only appear in the off-diagonal elements, but disappear in the diagonal elements of density matrix, so it will affect the density matrix , the entanglement entropy and the physical observable $< \mathrm{A} >= \mathrm{Tr}(\rho\hat{A})$.

In 2024, we further extended the above sub-geometric phase from adiabatic case to the non-adiabatic and non-cyclic case[14], which is similar to Samuel and Bhandari's phase[12] but with more sub-geometric phases correspond to each state. Both the real and imaginary parts of sub-geometric phase have influence on the quantum transition of the system in which a time dependent external field is applied. The imaginary part of sub-geometric phase will deviate the usual resonance peak of quantum transition, and affect the energy level crossing, while the real part of sub-geometric phase may determine the stability of initial state of the system, it will bring a useful complement on the selection rule of quantum transition[14].

In this manuscript, we will calculate the non-adiabatic sub-geometric phase in Kitaev QSL for a spin excitation state from the degenerate ground state, and investigate the stability of ground state by the real part of sub-geometric phase based on the linear stability analysis theory[15] which is used not only in classical non-equilibrium phase transition[16], but also in the robustness of quantum systems[17]. We also discuss the modification of the imaginary part of sub-geometric phase on the resonance peak of quantum transition. Finally, we will evaluate the entanglement entropy and spin correlation function, and explore the influence of sub-geometric phase on them, then extend the sub-geometric phase to the finite temperature case in Kitaev model.

## II. Theoretical Formalism

The Kitaev Hamiltonian for the spin system with honeycomb lattice is[5]

$$H_0 = J_x \sum_{<ij>\epsilon x} \sigma_i^x \sigma_j^x + J_y \sum_{<ij>\epsilon y} \sigma_i^y \sigma_j^y + J_x \sum_{<ij>\epsilon x} \sigma_i^z \sigma_j^z, \tag{1}$$

where $i, j$ denote the sites of the lattice, $< ij > \epsilon\alpha, \alpha = x, y, z$ labels the nearest neighbour bond in the α-th direction, $J_x$, $J_y$, $J_x$ are the exchange coupling constants in the $x, y, z$ directions, $\sigma_i^\alpha$ is the α-component of local spin operator at site $i$.

Define the plaquette operator $w_p = \sigma_1^x \sigma_2^y \sigma_3^z \sigma_4^x \sigma_5^y \sigma_6^z$, which commutes with the Hamiltonian $[w_p, H] = 0$ for all the plaquettes and $w_p = \pm 1$. Since the ground state have no vortices, $\forall w_p = +1$ for each plaquette in the ground state, then the ground state can be formally written as

$$|0 >= \prod_p Q_p (\otimes_i | \sigma_i^z = s_i), \tag{2}$$

where $s_i$=$\pm 1$ and $Q_p = \frac{1+w_p}{2}$ is the projector. The formal expression of Eq.(2) is not helpful for us to calculate the sub-geometric phase for an excited state. We need write exactly out each spin state in the ground state $|0>$. Since $\forall w_p = +1$ in the ground state, let us start with the state in which each spin in the $p$-plaquette $\{\sigma_{1p}^x = +1, \sigma_{2p}^y = +1, \sigma_{3p}^z = +1, \sigma_{4p}^x = +1, \sigma_{5p}^y = +1, \sigma_{6p}^z = +1\}$, if we want to keep $\{w_p = +1\}$ for all the plaquettes, we must flip all the six spins in the $p$-th plaquette to $\{\sigma_{1p}^x = -1, \sigma_{2p}^y = -1, \sigma_{3p}^z = -1, \sigma_{4p}^x = -1, \sigma_{5p}^y = -1, \sigma_{6p}^z = -1\}$ in order to assure all the $\{w_p = +1\}$, otherwise if we only flip two spins in one bond in the $p$-th plaquette, i.e., $\sigma_{1p}^x = -1, \sigma_{2p}^y = -1$, although $w_p$ is still equal to 1 for the $p$-th plaquette, however, $\sigma_{1p}^x$ is also belong to the neighbour plaquette near $p$, in which $w_p = -1$, which breaks the demand of all $\{w_p = +1\}$ in the ground state. Similarly, we can not only flip the spins in two, three ...or five bonds in plaquette $p$ simultaneously, which will lead to $w_p = -1$ in the neighbour plaquette, we must flip all the spins in six bonds of plaquette $p$ in order to keep $w_p = +1$ for all the plaquettes, so the ground state can be written as

$$|0>= \psi_{00}^0 + \psi_{\{p\}0}^1 + \psi_{\{p,q\}0}^2 + \psi_{\{p,q,r\}0}^3 + \ldots + \psi_{\{p,q\}0}^{N-2} + \psi_{\{p\}0}^{N-1} + \psi_{00}^N, \qquad (3)$$

where $\{\psi_{\{p,q,r\ldots\}0}^i\}$ label the spin degenerate states in the ground state, in which the spins of $i$ plaquettes which label by $\{p, q, r\ldots\}$ are completely flipped from $\{\sigma_{1p}^x = +1, \sigma_{2p}^y = +1, \sigma_{3p}^z = +1, \sigma_{4p}^x = +1, \sigma_{5p}^y = +1, \sigma_{6p}^z = +1\}$ to $\{\sigma_{1p}^x = -1, \sigma_{2p}^y = -1, \sigma_{3p}^z = -1, \sigma_{4p}^x = -1, \sigma_{5p}^y = -1, \sigma_{6p}^z = -1\}$, the subscript 0 denote the ground state. For brevity, we present the expressions of each spin state $\{\psi_{\{p,q,r\ldots\}0}^i\}$ in the ground state in appendix A. If the system have $N$ plaquettes, the number of spin states in $\psi_{00}^0$ is $C_N^0$, in $\psi_{\{p\}0}^1$ is $C_N^1$, ... , and in $\{\psi_{\{p,q,r\ldots\}0}^i\}$ is $C_N^i$, so the number of total spin states in the ground states is $C_N^0 + C_N^1 + C_N^2 + \ldots + C_N^{N-2} + C_N^{N-1} + C_N^N = 2^N$, they are highly degenerate.

If we apply a time dependent perturbation to the system, i.e., we only flip the 3-th spin in the $i$-th plaquette, the perturbation in Hamiltonian (1) can be written as

$$H' = B(t)\sigma_{1i}^{x}\sigma_{2i}^{y}\sigma_{3i}^{x}\sigma_{4i}^{x}\sigma_{5i}^{y}\sigma_{6i}^{z}, \tag{4}$$

where $\sigma_{3i}^{x}|\uparrow>=|\downarrow>$ and $\sigma_{3i}^{x}|\downarrow>=|\uparrow>$, it flips the 3-th spin states of $\sigma_{3i}^{z}$ in the $i$-th plaquette, then the ground state will be excited to a spin state with higher energies, it can be obtained by the time dependent perturbation theory as

$$|\psi(\mathrm{t})>=$$

$$\mathrm{c}_0^0(t)e^{-\frac{iE_0^0 t}{\hbar}}|\psi_0^0>+\mathrm{c}_0^1(t)e^{-\frac{iE_0^1 t}{\hbar}}|\psi_0^1>+\sum_{p\neq i}\mathrm{c}_\mathrm{p}^1(t)e^{-\frac{iE_p^1 t}{\hbar}}|\psi_p^1>+$$

$$\mathrm{c}_0^2(t)e^{-\frac{iE_0^2 t}{\hbar}}|\psi_0^2>+\sum_{p,q}\mathrm{c}_\mathrm{pq}^2(t)e^{-\frac{iE_{pq}^2 t}{\hbar}}|\psi_{pq}^2>$$

$$+\ldots+\mathrm{c}_0^{\mathrm{N}-2}(t)e^{-\frac{iE_0^{N-2} t}{\hbar}}|\psi_0^{N-2}>+\sum_{p,q}\mathrm{c}_\mathrm{pq}^{\mathrm{N}-2}(t)e^{-\frac{iE_{pq}^{N-2} t}{\hbar}}|\psi_{pq}^{N-2}>$$

$$+\mathrm{c}_0^{\mathrm{N}-1}(t)e^{-\frac{iE_0^{N-1} t}{\hbar}}|\psi_0^{N-1}>+\sum_{p\neq i}\mathrm{c}_\mathrm{p}^{\mathrm{N}-1}(t)e^{-\frac{iE_p^{N-1} t}{\hbar}}|\psi_p^1>$$

$$+\mathrm{c}_0^{\mathrm{N}}(t)e^{-\frac{iE_0^{N} t}{\hbar}}|\psi_0^N>, \tag{5}$$

where $|\psi_0^0>$, $|\psi_p^1>$, $|\psi_{pq}^2>$, ..., $|\psi_{pq}^{N-2}>$, $|\psi_p^{N-1}>$, $|\psi_0^N>$ denote the spin states flipping the 3-th spin of $i$-th plaquette from the spin states $|\psi_{00}^0>$, $|\psi_{p0}^1>$, $|\psi_{pq0}^2>$, ..., $|\psi_{pq0}^{N-2}>$, $|\psi_{p0}^{N-1}>$, $|\psi_{00}^N>$ in the ground state, respectively, while $|\psi_0^1>$, $|\psi_0^2>$, $|\psi_0^3>$, ..., $|\psi_0^{N-2}>$, $|\psi_0^{N-1}>$ label the spin states flipping the 3-th spin of $i$-th plaquette $(p = i)$ from the spin states $|\psi_{p0}^1>$, $|\psi_{pq0}^2>$, $|\psi_{pqr0}^3>$, ..., $|\psi_{pq0}^{N-2}>$, $|\psi_{p0}^{N-1}>$ in the ground state, respectively. For conciseness, the detailed expressions for the spin states

$|\psi_0^0>$, $|\psi_p^1>$, $|\psi_{pq}^2>$, ..., $|\psi_{pq}^{N-2}>$, $|\psi_p^{N-1}>$, $|\psi_0^N>$ and $|\psi_0^1>$, $|\psi_0^2>$, $|\psi_0^3>$, ..., $|\psi_0^{N-2}>$, $|\psi_0^{N-1}>$ are presented in appendix B. The eigen-energies in the above expression are written as: $\{E_{\{pqr\ldots\}}^i =< \psi_{\{pqr\ldots\}}^i | H_0 | \psi_{\{pqr\ldots\}}^i >, \mathrm{i} = 0{,}1 \ldots \mathrm{N}\}$, and the coefficients in expression (5) can be evaluated by the time dependent perturbation theory, they are :

$$\{c_{\{pqr\ldots\}}^i(t) = \frac{1}{i\hbar}\int_0^t e^{i(E_{\{pqr\ldots\}}^i - E_{\{pqr\ldots 0\}}^i)t'/\hbar} < \psi_{\{pqr\ldots\}}^i | H' | \psi_{\{pqr\ldots\}0}^i > dt', i = 0{,}1 \ldots N\}. \tag{6}$$

Let us choose an arbitrary coefficient from the above $\{c_{\{pqr\ldots\}}^i(t)\}$, which is a complex function and can be expressed as

$$c_{\{pqr\ldots\}}^i(t) = A_{\{pqr\ldots\}}^i(t) e^{i\varphi_{\{pqr\ldots\}}^i(t)}, \tag{7}$$

where $A_{\{pqr\ldots\}}^i(t)$ and $\varphi_{\{pqr\ldots\}}^i(t)$ are the modul and argument of the complex function $c_{\{pqr\ldots\}}^i(t)$, respectively. If we write $A_{\{pqr\ldots\}}^i(t) = e^{a_{\{pqr\ldots\}}^i(t)}$, where $a_{\{pqr\ldots\}}^i(t) = lnA_{\{pqr\ldots\}}^i(t)$, then $c_{\{pqr\ldots\}}^i(t)$ can be formally expressed as

$$c_{\{pqr\ldots\}}^i(t) = e^{a_{\{pqr\ldots\}}^i(t) + i\varphi_{\{pqr\ldots\}}^i(t)}, \tag{8}$$

where $a_{\{pqr\ldots\}}^i(t) + i\varphi_{\{pqr\ldots\}}^i(t)$ is just the sub-geometric phase given by us in Ref.[14]. We have given it's geometrical interpretation by the fibre bundle theory[14]. $a_{\{pqr\ldots\}}^i(t)$ and $\varphi_{\{pqr\ldots\}}^i(t)$ are the real and imaginary parts of sub-geometric phase, respectively, they are the generalization of non-adiabatic and non-cyclic Samuel and Bhandari's phase[12]. In the expression (5) of wavefunction for the excited state, we can combine the imaginary part of sub-geometric phase $e^{i\varphi_{\{pqr\ldots\}}^i(t)}$ together with the dynamical phase $e^{-iE^{i}_{\{pqr...\}}t/\hbar}$ in each term of Eq.(5), that is $e^{i(\varphi^{i}_{\{pqr...\}}(t)-E^{i}_{\{pqr...\}}t/\hbar)}$, the sub-geometric phase $\varphi^{i}_{\{pqr...\}}(t)$ has a modification on the eigen-energy $E^{i}_{\{pqr...\}}$ of QSL, which will leads to the deviation of resonance peak from $(E^{i}_{\{pqr...\}}-E^{i}_{\{p'q'r'...\}})/\hbar$ by an amount to $(E^{i}_{\{pqr...\}}-E^{i}_{\{p'q'r'...\}})/\hbar+(\frac{\hbar\varphi^{i}_{\{p'q'r'...\}}(t)}{t}-\frac{\hbar\varphi^{i}_{\{pqr...\}}(t)}{t})/\hbar$ when we apply a periodical perturbation with frequency $\omega=(E^{i}_{\{pqr...\}}-E^{i}_{\{p'q'r'...\}})/\hbar$ according to Wu's analysis[18]. So the imaginary part of sub-geometric phase will affect the level crossing during the quantum transition.

The real part of sub-geometric phase $a^{i}_{\{pqr...\}}(t)$ also play an important role in the quantum transition. If $a^{i}_{\{pqr...\}}(t)$ increases with time $t$, the initial state $\psi^{i}_{pqr...}$ will lose it's stability according to the linear stability analysis theory[15], and the quantum transition will occur. On the contrary, if $a^{i}_{\{pqr...\}}(t)$ decreases with time $t$, the initial state $\psi^{i}_{pqr...}$ will keep it's stability, the quantum transition will not happen, so even the transition element $<\psi^{i}_{pqr...}|\mathrm{H}'|\psi^{i}_{pqr...0}>$ is not zero, the quantum transition may not occur. Since $<\psi^{i}_{pqr...}|\mathrm{H}'|\psi^{i}_{pqr...0}>=0$ will lead to the forbidden of quantum transition and the selection rule, then the real part of sub-geometric phase $a^{i}_{\{pqr...\}}(t)$ will bring somewhat complement on the selection rule of quantum transition.

In the next, we will expound them by an example. If we choose $B(t)$ in perturbation (4) as

$$B(t) = Dexp[-i\omega t] \tag{9}$$

we can obtain the detailed expressions for the coefficients from Eq.(6). For example,

$$c_0^N(t) = \frac{1}{i\hbar}\int_0^t e^{i(E_0^N - E_{00}^N)t'/\hbar} Dexp[-i\omega t']dt' = a + ib, \tag{10}$$

where $a = \frac{D}{\hbar(\omega_0^N - \omega)}(1 - cos((\omega_0^N - \omega)\mathrm{t}))$ and $b = -\frac{D}{\hbar(\omega_0^N - \omega)} sin((\omega_0^N - \omega)\mathrm{t})$, and $\omega_0^N = (E_0^N - E_{00}^N)/\hbar$. If we write $c_0^N(t) = A_0^N(t)e^{i\varphi_0^N(t)}$, where $A_0^N(t) = D\sqrt{\frac{2}{\hbar^2(\omega_0^N - \omega)^2}(1 - cos((\omega_0^N - \omega)\mathrm{t}))}$ and $\varphi_0^N(t) = arctan\frac{b}{a}$, they are just the real and imaginary parts of sub-geometric phase. When $2k\pi \leq (\omega_0^N - \omega)\mathrm{t} \leq (2k+1)\pi$, $A_0^N(t)$=$e^{a_0^N(t)}$ increases with time $t$, the initial state $\psi_0^N$ lose it's stability and the quantum transition will occur. When $(2k+1)\pi \leq (\omega_0^N - \omega)\mathrm{t} \leq (2k+2)\pi$, $a_0^N(t)$ decreases with time $t$, the initial state will keep it's stability, the quantum transition will not happen. While the imaginary part $\varphi_0^N(t)$ of sub-geometric phase will bring modification on the resonant energy level from $E_0^N$ to $E_0^N - \frac{\hbar\varphi_0^N(t)}{t}$.

## III. The Entanglement Entropy and Spin Correlation Function

The sub-geometric phase also have effect on the entropy and other physical observable. From expression (5), the density matrix $\hat{\rho} = |\psi(\mathrm{t})\rangle\langle\psi(\mathrm{t})|$ is a $2^N \times 2^N$ dimensional matrix, which is shown in Appendix C. We can see that it is affected by the sub-geometric phase which doesn't appear in the diagonal elements, but appear in it's off-diagonal elements as a difference of sub-geometric phases corresponding to the two states of off-diagonal element. Substituting the density matrix into Von-Neumann entanglememt entropy $S_A = Tr_B \hat{\rho} ln \hat{\rho}$, where A denote the A sub-lattice in the Kagome lattice, while B label the B sub-lattice in this system, then the sub-geometric phase also has influence on the entanglement entropy. Similarly, for an arbitrary physical operator $\hat{A}$, the corresponding physical observable $< \mathrm{A} > = \mathrm{Tr}(\hat{\rho}\hat{A})$ is also affected by the sub-geometric phase through the density matrix.

Finally, we discuss the spin correlation function in QSL, which is defined as $\sigma_{ij}^{\alpha\beta} =< \psi(\mathrm{t})|\sigma_i^{\alpha}(t)\sigma_j^{\beta}(0)|\psi(\mathrm{t})>$. According to the study by Baskaran et al.[19], only the nearest neighbor $\mathrm{i}, \mathrm{j}$ and the same $\alpha, \beta$ have contribution to this function, only $\sigma_{<ij>}^{\alpha\alpha}$ $(\alpha = x, y, z)$ are not zero, they are

$$\sigma_{<ij>}^{\alpha\alpha} = c_0^{0*}(t)c_0^0(0)e^{\frac{iE_0^0 t}{\hbar}} + c_1^{1*}(t)c_1^1(0)e^{\frac{iE_1^1 t}{\hbar}} + \cdots + c_0^{N*}(t)c_0^N(0)e^{\frac{iE_0^N t}{\hbar}} = A_0^0(t)A_0^0(0)e^{-i\left(\varphi_0^0(t)-\varphi_0^0(0)-\frac{E_0^0 t}{\hbar}\right)} + A_1^1(t)A_1^1(0)e^{-i\left(\varphi_1^1(t)-\varphi_1^1(0)-\frac{E_1^1 t}{\hbar}\right)} + \cdots + A_0^N(t)A_0^N(0)e^{-i\left(\varphi_0^N(t)-\varphi_0^N(0)-\frac{E_0^N t}{\hbar}\right)}. \quad (11)$$

So the sub-geometric phases also have effect on the spin correlation function. Since the spin correlation function can be observed by neutron scattering in QSL, then the sub-geometric phase can be used to characterize the topological properties of QSL.

## IV. The Sub-Geometric Phase in the Finite Temperature Case

For the finite temperature case, the thermal state is not a pure state but a mixed state. Bascone et al. ever extended the Berry curvature of pure state to the mean Uhlmann curvature for the finite-temperature Kitaev model[10], the evolution of state in parameter space is still adiabatic. For our non-adiabatic and non-cyclic sub-geometric phase, the generalization to finite temperature case is straightforward. To a mixed state at temperature T, if we denote it's density matrix as $\rho = \sum_k p_k \rho_k$, where $\rho_k$ is the density matrix of the pure state $|\psi_k>$ $(k = 1,2 \ldots)$ and $p_k$ is the probability for $|\psi_k>$ appearing in the mixed state, i.e. the Boltzmann distribution or the Fermi distribution et al., there is no fixed relative phase including the relative geometric phase between different pure states. Since the sub-geometric phases still exist in the pure state $|\psi_k>$ $(k = 1,2 \ldots)$ or it's density matrix as shown in C2, substituting Eq.C2 into the density matrix $\rho = \sum_k p_k \rho_k$ of mixed thermal state, we naturally extend our sub-geometric phase to the finite temperature case. Despite some of the treatments proposed[20-22] for the geometric phase in a mixed state, defining the geometric phase in a mixed state is still difficult. However, in the above we present an alternative reasonable treatment.

## V. Summary and Discussions

Based the sub-geometric phase proposed by us[14], we calculate the corresponding phase in Kitaev QSL. We give the detailed expression for the many body wavefunction of degenerate ground state and a kind of excited spin state, and demonstrate the sub-geometric phase therein. These sub-geometric phase can naturally appear in the density matrix, entanglement entropy, spin correlation function as well as other physical observable, so it have effect on them. Since the spin correlation function can be observed by neutron scattering in QSL, then the sub-geometric phase can be used as a physical quantity to characterize the topological properties of QSL. Our sub-geometric phase can be regarded as a generalization of non-cyclic and non-adaibatic Samuel and Bhandari's phase, but the latter can not appear in the density matrix and other physical observable. So our generalization on the sub-geometric phase is necessary for QSL. We show that the imaginary part of sub-geometric phase will deviate the resonance peak of quantum transition and modify the energy level crossing, while it's real part determine the stability of initial state, it provide a convenient complement on the selection rule of quantum transition. We also extend our sub-geometric phase to the finite temperature case. Certainly, the time dependent perturbation Eq.(4) in the Hamiltonian chosen in our manuscript is very simple, we only flip one spin in a plaquette, the more general perturbation can be discussed in the future exploration.

## Acknowledgments

This study is supported by the National Key R&D Program of China (Grant No. 2022YFA1402703), the Strategic Priority Research Program of the Chinese Academy of Sciences (Grant No. XDB28000000).

## Appendix A

The eigen-states in the ground state (3) are expressed as follows:

$\psi_{00}^{0} = \prod_p \otimes(|\sigma_{1p}^{x} = +1, \sigma_{2p}^{y} = +1, \sigma_{3p}^{z} = +1, \sigma_{4p}^{x} = +1, \sigma_{5p}^{y} = +1, \sigma_{6p}^{z} = +1 >)$ , A1

$\psi_{\{p\}0}^{1} = \sum_p |\sigma_{1p}^{x} = -1, \sigma_{2p}^{y} = -1, \sigma_{3p}^{z} = -1, \sigma_{4p}^{x} = -1, \sigma_{5p}^{y} = -1, \sigma_{6p}^{z} = -1 > \prod_l \otimes(|\sigma_{1l}^{x} = +1, \sigma_{2l}^{y} = +1, \sigma_{3l}^{z} = +1, \sigma_{4l}^{x} = +1, \sigma_{5l}^{y} = +1, \sigma_{6l}^{z} = +1 > \{l \neq p\})$, A2

$\psi_{\{p,q\}0}^{2} = \sum_{p,q} |\sigma_{1p}^{x} = -1, \sigma_{2p}^{y} = -1, \sigma_{3p}^{z} = -1, \sigma_{4p}^{x} = -1, \sigma_{5p}^{y} = -1, \sigma_{6p}^{z} = -1 > \otimes |\sigma_{1q}^{x} = -1, \sigma_{2q}^{y} = -1, \sigma_{3q}^{z} = -1, \sigma_{4q}^{x} = -1, \sigma_{5q}^{y} = -1, \sigma_{6q}^{z} = -1 > \prod_l \otimes(|\sigma_{1l}^{x} = +1, \sigma_{2l}^{y} = +1, \sigma_{3l}^{z} = +1, \sigma_{4l}^{x} = +1, \sigma_{5l}^{y} = +1, \sigma_{6l}^{z} = +1 > \{l \neq p, q\})$ , A3

...

$\psi_{\{p,q\}0}^{N-2} = \sum_{p,q} |\sigma_{1p}^{x} = +1, \sigma_{2p}^{y} = +1, \sigma_{3p}^{z} = +1, \sigma_{4p}^{x} = +1, \sigma_{5p}^{y} = +1, \sigma_{6p}^{z} = +1 > \otimes |\sigma_{1q}^{x} = +1, \sigma_{2q}^{y} = +1, \sigma_{3q}^{z} = +1, \sigma_{4q}^{x} = +1, \sigma_{5q}^{y} = +1, \sigma_{6q}^{z} = +1 > \prod_l \otimes(|\sigma_{1l}^{x} = -1, \sigma_{2l}^{y} = -1, \sigma_{3l}^{z} = -1, \sigma_{4l}^{x} = -1, \sigma_{5l}^{y} = -1, \sigma_{6l}^{z} = -1 > \{l \neq p, q\})$ , A4

$\psi_{\{p\}0}^{N-1} = \sum_p |\sigma_{1p}^{x} = +1, \sigma_{2p}^{y} = +1, \sigma_{3p}^{z} = +1, \sigma_{4p}^{x} = +1, \sigma_{5p}^{y} = +1, \sigma_{6p}^{z} =$

$+1 > \prod_l \otimes(|\sigma_{1l}^x = -1, \sigma_{2l}^y = -1, \sigma_{3l}^z = -1, \sigma_{4l}^x = -1, \sigma_{5l}^y = -1, \sigma_{6l}^z = -1 > \{l \neq p\})$, A5

$\psi_{00}^N = \prod_p \otimes(|\sigma_{1p}^x = -1, \sigma_{2p}^y = -1, \sigma_{3p}^z = -1, \sigma_{4p}^x = -1, \sigma_{5p}^y = -1, \sigma_{6p}^z = -1 >)$. A6

## Appendix B

The eigen-states in the excited state (5) are expressed as follows:

$|\psi_0^0 >= |\sigma_{1i}^x = +1, \sigma_{2i}^y = +1, \sigma_{3i}^z = -1, \sigma_{4i}^x = +1, \sigma_{5i}^y = +1, \sigma_{6i}^z = +1 > \prod_{l\neq i} \otimes(|\sigma_{1l}^x = +1, \sigma_{2l}^y = +1, \sigma_{3l}^z = +1, \sigma_{4l}^x = +1, \sigma_{5l}^y = +1, \sigma_{6l}^z = +1 >)$, B1

$|\psi_0^1 >= |\sigma_{1i}^x = -1, \sigma_{2i}^y = -1, \sigma_{3i}^z = +1, \sigma_{4i}^x = -1, \sigma_{5i}^y = -1, \sigma_{6i}^z = -1 > \prod_{l\neq i} \otimes(|\sigma_{1l}^x = +1, \sigma_{2l}^y = +1, \sigma_{3l}^z = +1, \sigma_{4l}^x = +1, \sigma_{5l}^y = +1, \sigma_{6l}^z = +1 >)$, B2

$|\psi_p^1 >= \sum_p |\sigma_{1p}^x = -1, \sigma_{2p}^y = -1, \sigma_{3p}^z = -1, \sigma_{4p}^x = -1, \sigma_{5p}^y = -1, \sigma_{6p}^z = -1 > \otimes |\sigma_{1i}^x = +1, \sigma_{2i}^y = +1, \sigma_{3i}^z = -1, \sigma_{4i}^x = +1, \sigma_{5i}^y = +1, \sigma_{6i}^z = +1 > \prod_{l\neq i} \otimes(|\sigma_{1l}^x = +1, \sigma_{2l}^y = +1, \sigma_{3l}^z = +1, \sigma_{4l}^x = +1, \sigma_{5l}^y = +1, \sigma_{6l}^z = +1 >)$, B3

$|\psi_0^2 >= |\sigma_{1i}^x = -1, \sigma_{2i}^y = -1, \sigma_{3i}^z = +1, \sigma_{4i}^x = -1, \sigma_{5i}^y = -1, \sigma_{6i}^z = -1 > \otimes \sum_{p\neq i} |\sigma_{1p}^x = -1, \sigma_{2p}^y = -1, \sigma_{3p}^z = -1, \sigma_{4p}^x = -1, \sigma_{5p}^y = -1, \sigma_{6p}^z = -1 > \prod_{l\neq i,p} \otimes(|\sigma_{1l}^x = +1, \sigma_{2l}^y = +1, \sigma_{3l}^z = +1, \sigma_{4l}^x = +1, \sigma_{5l}^y = +1, \sigma_{6l}^z = +1 >)$, B4

$|\psi_{pq}^2 >= |\sigma_{1p}^x = -1, \sigma_{2p}^y = -1, \sigma_{3p}^z = -1, \sigma_{4p}^x = -1, \sigma_{5p}^y = -1, \sigma_{6p}^z = -1 > \otimes |\sigma_{1q}^x = -1, \sigma_{2q}^y = -1, \sigma_{3q}^z = -1, \sigma_{4q}^x = -1, \sigma_{5q}^y = -1, \sigma_{6q}^z =$

$-1 > \otimes|\sigma_{1i}^x = +1, \sigma_{2i}^y = +1, \sigma_{3i}^z = -1, \sigma_{4i}^x = +1, \sigma_{5i}^y = +1, \sigma_{6i}^z =$
$+1 > \prod_{l\neq p,q,i} \otimes(|\sigma_{1l}^x = +1, \sigma_{2l}^y = +1, \sigma_{3l}^z = +1, \sigma_{4l}^x = +1, \sigma_{5l}^y =$
$+1, \sigma_{6l}^z = +1 >),$ B5

…

$|\psi_0^{N-2} >= |\sigma_{1i}^x = +1, \sigma_{2i}^y = +1, \sigma_{3i}^z = -1, \sigma_{4i}^x = +1, \sigma_{5i}^y = +1, \sigma_{6i}^z =$
$+1 > \otimes \sum_{p\neq i} |\sigma_{1p}^x = +1, \sigma_{2p}^y = +1, \sigma_{3p}^z = +1, \sigma_{4p}^x = +1, \sigma_{5p}^y =$
$+1, \sigma_{6p}^z = +1 > \prod_{l\neq i,p} \otimes(|\sigma_{1l}^x = -1, \sigma_{2l}^y = -1, \sigma_{3l}^z = -1, \sigma_{4l}^x =$
$-1, \sigma_{5l}^y = -1, \sigma_{6l}^z = -1 >),$ B6

$|\psi_{pq}^{N-2} >= |\sigma_{1p}^x = +1, \sigma_{2p}^y = +1, \sigma_{3p}^z = +1, \sigma_{4p}^x = +1, \sigma_{5p}^y = +1, \sigma_{6p}^z =$
$+1 > \otimes|\sigma_{1q}^x = +1, \sigma_{2q}^y = +1, \sigma_{3q}^z = +1, \sigma_{4q}^x = +1, \sigma_{5q}^y = +1, \sigma_{6q}^z =$
$+1 > \otimes|\sigma_{1i}^x = -1, \sigma_{2i}^y = -1, \sigma_{3i}^z = +1, \sigma_{4i}^x = -1, \sigma_{5i}^y = -1, \sigma_{6i}^z =$
$-1 > \prod_{l\neq p,q,i} \otimes(|\sigma_{1l}^x = -1, \sigma_{2l}^y = -1, \sigma_{3l}^z = -1, \sigma_{4l}^x = -1, \sigma_{5l}^y =$
$-1, \sigma_{6l}^z = -1 >),$ B7

$|\psi_0^{N-1} >= |\sigma_{1i}^x = +1, \sigma_{2i}^y = +1, \sigma_{3i}^z = -1, \sigma_{4i}^x = +1, \sigma_{5i}^y = +1, \sigma_{6i}^z =$
$+1 > \prod_{l\neq i} \otimes(|\sigma_{1l}^x = -1, \sigma_{2l}^y = -1, \sigma_{3l}^z = -1, \sigma_{4l}^x = -1, \sigma_{5l}^y =$
$-1, \sigma_{6l}^z = -1 >),$ B8

$|\psi_p^{N-1} >= \sum_{p\neq i} |\sigma_{1p}^x = +1, \sigma_{2p}^y = +1, \sigma_{3p}^z = +1, \sigma_{4p}^x = +1, \sigma_{5p}^y =$
$+1, \sigma_{6p}^z = +1 > \otimes|\sigma_{1i}^x = -1, \sigma_{2i}^y = -1, \sigma_{3i}^z = +1, \sigma_{4i}^x = -1, \sigma_{5i}^y =$
$-1, \sigma_{6i}^z = -1 > \prod_{l\neq i} \otimes(|\sigma_{1l}^x = -1, \sigma_{2l}^y = -1, \sigma_{3l}^z = -1, \sigma_{4l}^x =$
$-1, \sigma_{5l}^y = -1, \sigma_{6l}^z = -1 >),$ B9

$|\psi_0^N >= |\sigma_{1i}^x = -1, \sigma_{2i}^y = -1, \sigma_{3i}^z = +1, \sigma_{4i}^x = -1, \sigma_{5i}^y = -1, \sigma_{6i}^z = -1 >$
$\prod_{l\neq i} \otimes(|\sigma_{1l}^x = -1, \sigma_{2l}^y = -1, \sigma_{3l}^z = -1, \sigma_{4l}^x = -1, \sigma_{5l}^y = -1, \sigma_{6l}^z =$

$-1 >)$.  B10

## Appendix C

The density operator is written as:

$$\hat{\rho} = |\psi(\mathrm{t})>< \psi(\mathrm{t})|, \quad \text{C1}$$

Substituting the expression (5) of $|\psi(\mathrm{t})>$ into the above equation, the density matrix can further be expressed as:

$$\hat{\rho} = c_0^{0*}(t)c_0^0(t)|\psi_0^0 >< \psi_0^0| + c_0^{1*}(t)c_0^0(t)e^{i(E_0^1-E_0^0)t\prime/\hbar}|\psi_0^0 >< \psi_0^1| + \sum_{p\neq i} c_p^{1*}(t)c_0^0(t)e^{i(E_p^1-E_0^0)t\prime/\hbar}|\psi_0^0 >< \psi_p^1| + \ldots + c_0^{N-1*}(t)c_0^0(t)e^{i(E_0^{N-1}-E_0^0)t\prime/\hbar}|\psi_0^0 >< \psi_0^{N-1}| + \sum_{\mathrm{p}\neq\mathrm{i}} c_p^{N-1*}(t)c_0^0(t)e^{i(E_p^{N-1}-E_0^0)t\prime/\hbar}|\psi_0^0 >< \psi_p^{N-1}| + c_0^{N*}(t)c_0^0(t)e^{i(E_0^N-E_0^0)t\prime/\hbar}|\psi_0^0 >< \psi_0^N| + c_0^{0*}(t)c_0^1(t)e^{i(E_0^0-E_0^1)t\prime/\hbar}|\psi_0^1 >< \psi_0^0| + c_0^{1*}(t)c_0^1(t)|\psi_0^1 >< \psi_0^1| + \sum_{\mathrm{p}\neq\mathrm{i}} c_p^{1*}(t)c_0^1(t)e^{i(E_p^1-E_0^1)t\prime/\hbar}|\psi_0^1 >< \psi_p^1| + \ldots + c_0^{N-1*}(t)c_0^1(t)e^{i(E_0^{N-1}-E_0^1)t\prime/\hbar}|\psi_0^1 >< \psi_0^{N-1}| + \sum_{\mathrm{p}\neq\mathrm{i}} c_p^{N-1*}(t)c_0^1(t)e^{i(E_p^{N-1}-E_0^1)t\prime/\hbar}|\psi_0^1 >< \psi_p^{N-1}| + c_0^{N*}(t)c_0^1(t)e^{i(E_0^N-E_0^1)t\prime/\hbar}|\psi_0^1 >< \psi_0^N| + \ldots + c_0^{0*}(t)c_0^N(t)e^{i(E_0^0-E_0^N)t\prime/\hbar}|\psi_0^N >< \psi_0^0| + c_0^{1*}(t)c_0^N(t)e^{i(E_0^1-E_0^N)t\prime/\hbar}|\psi_0^N >< \psi_0^1| + \sum_{\mathrm{p}\neq\mathrm{i}} c_p^{1*}(t)c_0^N(t)e^{i(E_p^1-E_0^N)t\prime/\hbar}|\psi_0^N >< \psi_p^1| + \ldots + c_0^{N-1*}(t)c_0^N(t)e^{i(E_0^{N-1}-E_0^N)t\prime/\hbar}|\psi_0^N >< \psi_0^{N-1}| + \sum_{\mathrm{p}\neq\mathrm{i}} c_p^{N-1*}(t)c_0^N(t)e^{i(E_p^{N-1}-E_0^N)t\prime/\hbar}|\psi_0^N >< \psi_p^{N-1}| + c_0^{N*}(t)c_0^N(t)|\psi_0^N >< \psi_0^N|. \quad \text{C2}$$

## References

[1] Y. Zhou, K. Kanoda and T. K. Ng, Rev. Mod. Phys. 89, 025003 (2017).

[2] H. T. Diep, (ed.) *Frustrated Spin Systems* (World Scientific, 2013).

[3] D. Bergmani, J. Alicea, E. Gull, S. Trebst and L. Balents, Nat. Phys. 3, 487 ( 2007).

[4] A. Kitaev, Ann. Phys.321, 2 (2006).

[5] L. M. Duan, E. Demler and M. D. Lukin, Phys. Rev. Lett. 91, 090402 (2003).

[6] L. Savary, L. Balents, arXiv:1601.03742v1: cond-mat., 2016.

[7] Y. Kasahara, Nature 559, 227 (2018).

[8] A. Shapere and F. Wilczek, *Geometric Phase in Physics* (World Scientific, Singapore, 1989).

[9] F. Bascone, L. Leonforte, D. Valenti, B. Spagnolo and A. Carollo, arXiv:1905.04118v1: cond-mat., 2019.

[10] F. Bascone, L. Leonforte, D. Valenti, B. Spagnolo and A. Carollo, Phys. Rev. B99, 205155 (2019).

[11] Y. Aharonov and J. Anandan, Phys. Rev. Lett. **58**, 1593(1987).

[12] J. Samuel and R. Bhandari, Phys. Rev. Lett. 60, 2339 (1988).

[13] Z. C. Wang, Scientific Reports, 9, 13258 (2019).

[14] Z. C. Wang, arXiv:2405.10697: quan-phys., 2024..

[15] R. Bellman, *Stability theory of Differential equations*, McGraw-Hill,

New York 1953.

[16] D. H. Sattinger, *Topics in stability and Bifurcation theory*, Springer, Berlin, 1973.

[17] Z. C. Wang, Mod. Phys. Lett. B36, 2250037 (2022).

[18] Y. S. Wu and H. Z. Li, Phys. Rev. B 38, 11907 (1988).

[19] G. Baskaran, S. Mandal and R. Shankar,, Phys. Rev. Lett. 98, 247201 (2007).

[20] E. Sjoqvist, A. K. Pati, A. Ekert, J. S. Anandan, M. Ericsson, D. K. L. Oi and V. Vedral, Phys. Rev. Lett. **85**, 2845 (2000).

[21] J. Du, P. Zou, M. Shi, L. C. Kwek and J. W. Pan, Phys. Rev. Lett. 91, 100403 (2003).

[22] A. Uhlmann, Ann. Phys. 501, 63 (1989).